\documentclass{article}
\begin{document}

\begin{center}
\centerline{\large \bf Femtosecond laser as a tool for experimental study}
\centerline{\large \bf of time invariance violation in optics.}
\end{center}

\vspace{3 pt}
\centerline{\sl V.A.Kuz'menko\footnote{Electronic 
address: kuzmenko@triniti.ru}}

\vspace{5 pt}
\centerline{\small \it Troitsk Institute for Innovation and Fusion 
Research,}
\centerline{\small \it Troitsk, Moscow region, 142190, Russian 
Federation.}
\vspace{5 pt}
\begin{abstract}

        The experiments with weak collinear probe pulses for time 
 reversal noninvariance study in optics are proposed and discussed.

\vspace{5 pt}
\end{abstract}

\vspace{12 pt}

        The equivalence of forward and reversed processes is a 
fundamental concept in optics today. However, this is only the hypothesis 
which does not have any experimental proofs. In the nearest future this 
concept should be rejected. From one side, because of it gives very meager 
possibility to explain physical origin of great number of phenomena in 
nonlinear optics. Frequently such phenomena have good mathematical 
description on the base of Bloch equations. However, these equations do 
not have clear physical sense. The concept of interference of 
"wave packets" or coherent states of molecules is usually used for 
explanation of the origin of nonlinear phenomena. However, the concept of 
coherent states in itself does not have clear and reliable physical 
base [1,2]. The concept of inequality of forward and reversed processes 
is the alternative analog of the concept of coherent states. In contrast 
to latter concept, it has clear physical sense and gives good base for 
explanation of the origin of number of phenomena in nonlinear optics [3]. 

        From other side, the equivalence concept should be 
rejected because of the opposite concept even today has several direct 
experimental proofs [4]. However, the concept of time noninvariance is 
so radical for optics, that it is not clear today how many experimental 
proofs will need for its recognition. A femtosecond lasers should 
play important role here. 
        
        The obtained experimental results show, that the difference in 
cross-sections of forward and reversed processes can exceed many 
orders of magnitude [5, 6]. The cross-section of reversed transition, 
obviously, has very sharp dependence from the orientation of molecules 
in space, from the phase of vibrational motion of atoms and even from the 
phase of laser radiation in the space. The classical pump-probe experiments 
with femtosecond laser pulses will give, probably, the most exact 
information of such kind. In such experiments two demands must be fulfilled:

1) - the pump and probe beams should be exactly collinear,

2) - the pump pulse should not saturate optical transition and the probe 
pulse should be very weak avoiding changing the distribution of species on 
the energy levels. \\ 
Only such arrangement allows to obtain correctly information about the 
cross-section of reversed optical transitions. The principle set up of 
such experiments is shown in Fig.1a, where a fluorescence of excited 
species is used in a detection scheme. 

        It is surprising, but among the huge number of pump-probe 
experiments with femtosecond lasers such experimental arrangement is not 
used till now. The weak probe pulse is usually used in the cross beams 
geometry. This case suits for four photon mixing, but does not suit for 
the study of the reversed transitions into the initial state. 

        The most closely related experiments were carried out in [7-10]. 
The authors studied the fluorescence of molecules or atoms (which 
characterizes the population of excited states) after it interaction 
with the pair of collinear phase-locked femtosecond laser pulses. However, 
the authors used equal intensity of pump and probe pulses. Such experiments 
should be continued in low intensity regime with weak collinear probe 
pulse. Close to ideal intensity regime was used in [11], but the demand of 
colinearity was not fulfilled. In proper conditions some amplification of 
probe laser radiation without inversion should be observed in some degree 
similar to results of [5]. It allows to make the evaluation of difference 
in cross-sections of forward and reversed transitions. 

        Here, the widely spread mistake exists that femtosecond laser 
radiation can align molecules in the so-called field free regime [12, 13]. 
This point of view does not have reliable experimental proofs. Our 
explanation is that the observed revivals are the manifestation of very 
sharp dependence of the reversed Raman transition's cross-section from 
orientation of molecule in the space [14]. The discussed low intensity 
collinear pump-probe experiments will make this situation clear. The 
principle scheme for reversed Raman transitions study is shown in Fig.1b. 
As a detection tool, the high harmonic generation (HHG) [15], the Coulomb 
explosion imaging [16] or polarization [17] techniques   may be used. The 
pump femtosecond pulse produces forward Raman transitions between the 
ground states of molecules. The strong reading pulse is fixed in time and 
turned on the strong delayed revival. The weak collinear probe laser pulse 
is used for study the efficiency of reversed Raman transitions. Rather 
close experiments were carried out recently in [17, 18]. However, again 
the intensity of probe pulses was too strong.
 
        As a detection scheme in such experiments a four photon mixing 
phenomenon in the so-called boxcars arrangement may be used [19, 20]. 
In such way the high quality experimental result with Raman transitions 
were obtained recently in [21]. It demonstrates not the suppression of 
alignment, but the erasing of stored information by efficient reversed 
process under action of the probe pulse (as a whole quite similar to 
those in [10]). The additional experiments with low intensity collinear 
probe pulses will give some information about relative cross-section 
of the reversed Raman transitions into the initial state. 

        In conclusion, we discussed the experiments with weak collinear 
probe femtosecond laser pulses for time invariance violation study 
in optics. Such experiments will allow us to understand and to give 
clear physical explanation for number of nonlinear phenomena, 
like as the population transfer or so-called "coherent control" [22].


\vspace{5 pt}
     
     
\begin{thebibliography}{99}
\bibitem{1} K.Molmer, Phys.Rev.A {\bf 55}, 3195 (1997).
\bibitem{2} K.Nemoto, and S.L.Braunstein, E-print, quant-ph/0312108.
\bibitem{3} V.A.Kuz'menko, E-print, physics/0306148.
\bibitem{4} V.A.Kuz'menko, E-print, physics/0506023.
\bibitem{5} C.Liedenbaum, S.Stolte, and J.Reuss, Chem.Phys. 
{\bf 122}, 443 (1988).
\bibitem{6} B.Dayan, A.Pe'er, A.A.Friesem and Y.Silberberg, E-print, 
quant-ph/0401088.
\bibitem{7} N.F.Scherer, A.J.Ruggiero, M.Du, and G.R.Fleming, 
J.Chem.Phys. {\bf 93}, 856 (1990).
\bibitem{8} N.F.Scherer, R.J.Carlson, A.Matro, M.Du, A.J.Ruggiero, 
V.Romero-Rochin, J.A.Cina, G.R.Fleming, and S.A.Rice, 
J.Chem.Phys. {\bf 95}, 1487 (1991). 
\bibitem{9} H.Yamada, K.Yokoyama, Y.Teranishi, A.Sugita, T.Shirai, 
M.Aoyama, Y.Akahane, N.Inoue, H.Ueda, K.Yamakawa, A.Yokoyama, M.Kawasaki, 
and H.Nakamura, Phys.Rev. A {\bf 72}, 063404 (2005).
\bibitem{10} K.Ohmori, H.Katsuki, H.Chiba, M.Honda, Y.Hagihara, K.Fujiwara, 
Y.Sato, and K.Ueda, Phys.Rev.Lett. {\bf 96}, 093002 (2006).
\bibitem{11} V.I.Prokhorenko, A.M.Nagy, and R.J.D.Miller, 
J.Chem.Phys. {\bf 122}, 184502 (2005).
\bibitem{12} E.Peronne, M.D.Poulsen, C.Z.Bisgaard, H.Stapelfeldt, 
and T.Seideman, Phys.Rev.Lett. {\bf 91}, 043003 (2003).
\bibitem{13} I.V.Litvinyuk, K.F.Lee, P.W.Dooley, D.M.Rayner, D.M.Villeneuve, 
and P.B.Corkum, Phys.Rev.Lett. {\bf 90}, 233003 (2003). 
\bibitem{14} V.A.Kuz'menko, E-print, physics/0310090.
\bibitem{15} K.Miyazaki, M.Kaku, G.Miyaji, A.Abdurrouf, and F.H.M.Faisal, 
Phys.Rev.Lett. {\bf 95}, 243903 (2005). 
\bibitem{16} H.Stapelfeldt, and T.Seideman, Rev.Mod.Phys. 
{\bf 75}, 543 (2003).
\bibitem{17} M.Renard, E.Hertz, S.Guerin, H.R.Jauslin, B.Lavorel, 
and O.Faucher, Phys.Rev.A, {\bf 72}, 025401 (2005).
\bibitem{18} K.F.Lee, E.A.Shapiro, D.M.Villeneuve, and P.B.Corkum, 
Phys.Rev.A {\bf 73}, 033403 (2006).
\bibitem{19} Y.Prior, Appl.Opt. {\bf 19}, 1741 (1980).
\bibitem{20} M.Schmitt, G.Knopp, A.Materny, and W.Kiefer, 
Chem.Phys.Lett. {\bf 270}, 9 (1997).
\bibitem{21} Sh.Fleischer, I.Sh.Averbukh, and Y.Prior, 
E-print, quant-ph/0601197.
\bibitem{22} M.Dantus, and V.V.Lozovoy, Chem.Rev.  {\bf 104}, 1813 (2004).

\end{thebibliography}
\end{document}